%% file: phipi.tex
\documentclass[twocolumn,showpacs,aps,prl,superscriptaddress]{revtex4}

\usepackage{graphicx}
\usepackage{dcolumn}
\usepackage{amsmath}
\usepackage{epsfig}

\input pubboard/babarsym

\long\def\inst#1{\par\nobreak\kern 4pt\nobreak
    {\it #1}\par\vskip 10pt plus 3pt minus 3pt}

\newcommand{\BABARPubYear}    {06}
\newcommand{\BABARPubNumber}  {022}

\newcommand{\SLACPubNumber}   {11854}
\newcommand{\LANLNumber}      {0605037}

\def\figurebox#1#2#3{%
    \def\arg{#3}%
    \ifx\arg\empty
    {\hfill\vbox{\hsize#2\hrule\hbox to #2{\vrule\hfill\vbox to #1{\hsize#2\vfill}\vrule}\hrule}\hfill}%
    \else
    {\hfill\epsfbox{#3}\hfill}%
    \fi}

\begin{document}

\preprint{\babar-PUB-\BABARPubYear/\BABARPubNumber} 
\preprint{SLAC-PUB-\SLACPubNumber} 

\begin{flushleft}
\babar-PUB-\BABARPubYear/\BABARPubNumber\\
SLAC-PUB-\SLACPubNumber\\
hep-ex/\LANLNumber\\[10mm]
\end{flushleft}

\title{
\Large \bf Search for {\boldmath {\boldmath $\B^+\to\phi\pi^+$}\ and 
$\Bz\to\phi\piz$}\ Decays
}

\input pubboard/authors_feb2006.tex

\date{\today}

\begin{abstract}
 
A search has been made for the decays $\B^+\to\phi\pi^+$ and  $\Bz\to\phi\piz$ 
 in a data sample of approximately 232 $\times 10^6$
$B\bar B$ pairs recorded at the $\Upsilon(4S)$ resonance with the \babar\ detector 
at the \pep2\ $B$-meson Factory at SLAC. No significant
 signals have been observed, and therefore  upper limits have been set 
on the branching fractions:  
$\BR(\B^+\to\phi\pi^+)<2.4 \times   10^{-7}$ and $\BR(\Bz\to\phi\piz)<2.8 \times 10^{-7}$ at  
$90\%$ probability.
\end{abstract}


\maketitle

\label{sec:Introduction}
The measurements of $B \to \phi K$ and $B \to \phi \pi$ decay rates are
important because they are
sensitive to contributions beyond the Standard Model (SM). 
In particular, the latter is strongly suppressed in the SM,
and a measurement of ${\mathcal B}(B \to \phi \pi) \gtrsim 10^{-7}$
would be evidence for new physics, for example supersymmetric contributions~\cite{Bar-Shalom:2002sv}.
The study of the processes $B^+ \to\phi\pi^+$~\cite{charge}
and $B^0 \to\phi\pi^0$  is also
important to understand the theoretical uncertainties associated with 
measurements of \CP asymmetries in $\Bz \to \phi\Kz$ decays.
The $B \to \phi\pi$ decay amplitudes are related to 
the sub-leading terms of the $\Bz \to \phi\Kz$ decay amplitude~\cite{burassilv} and 
can therefore provide stringent bounds on possible 
contributions to the time-dependent \CP asymmetry in $B^0\to\phi K^0$~\cite{gqbound}, 
another probe of new physics effects in $B$ decays. 

\begin{figure}[!ht]
\begin{center}
\epsfig{file=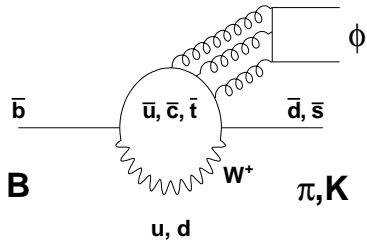,width=9cm}
\caption{Feynman diagram for $B \to \phi\pi$ and $B \to \phi K$.\label{fig:fey}}
\end{center}
\end{figure}

In Fig.~\ref{fig:fey} we show the leading order Feynman diagram for  the $B \to \phi \pi$ 
decay and a sub-leading diagram for $B \to \phi K$ decay.

Previous searches for these decay modes have been reported by \babar{} and CLEO~\cite{bfphik,previousbabar,cleo}.
The results presented here are based on data collected with the \babar\
detector~\cite{ref:babar} at the PEP-II asymmetric-energy $e^+e^-$
collider~\cite{pep} located at the Stanford Linear Accelerator Center.
An integrated luminosity of 211 \invfb, corresponding to
$(231.8 \pm 2.6) \times 10^6$ \BB\ pairs, was recorded at the
$\Upsilon (4S)$ resonance (center-of-mass energy $\sqrt{s}=10.58\
\gev$).

Charged particles from the \epem\ interactions are detected, and their
momenta measured, by a combination of five layers of double-sided
silicon microstrip detectors (SVT) and a 40-layer drift chamber (DCH),
both operating in the 1.5-T magnetic field of a superconducting
solenoid. Photons and electrons are identified with a CsI(Tl)
electromagnetic calorimeter (EMC).  Further charged particle
identification (PID) is provided by the average energy loss (\dedx) in
the tracking devices and by an internally reflecting ring imaging
Cherenkov detector (DIRC) covering the central region.

Monte Carlo simulation is used to 
evaluate background contamination and selection efficiency. Signal 
and background Monte Carlo samples are generated with EvtGen \cite{Lange:2001uf}. The detector response is 
simulated with GEANT4 \cite{geant} and all simulated events are reconstructed in 
the same manner as data.

We reconstruct $B$ meson candidates through the decays $\phi \pi^+$ or $\phi \pi^0$, with $\phi \to K^+K^-$ and
$\pi^0 \to \gamma\gamma$.
All kaon candidate tracks in the reconstructed decay chains must satisfy a set of 
loose kaon identification criteria based on the response of the DIRC and the 
\dedx measurements in the DCH and SVT. In both decay modes, all the tracks 
coming from the fully reconstructed $B$ are required to originate from the interaction point.
A pair of oppositely-charged kaon candidates is considered as a $\phi$ candidate if its 
invariant mass is within $15\mevcc$ of the nominal $\phi$ mass value (1019.5\mevcc \cite{pdg2004}). 
This is about three times the observed width in the $K^+K^-$ invariant mass spectrum.
A pair of energy deposits in the EMC, each of which is isolated from any charged
track and has the lateral shower shape expected for photons, is considered
as a $\pi^0$ candidate if both the deposits exceed $40\mev$ in the laboratory frame and the associated
invariant mass of the pair is between $110\mevcc$ and $160\mevcc$
(about three times the observed width in the $\gamma\gamma$ invariant mass spectrum).
$B$ meson candidates are made by 
combining $\phi$ candidates with a charged track or a $\pi^0$ candidate.
We do not apply any particle identification criteria on the track which comes directly from 
$B$ meson decay (primary track) at this stage,
so for the charged mode we reconstruct  $B^+ \to \phi h^+$ ($h^+ = \pi, K$) events.
This allows us to study the $B^+\to \phi K^+$ signal, which is the largest background 
coming from $B$ decays.

Two kinematic variables are used  to discriminate between signal $B$ 
decays and combinatorial background:
  the invariant mass of the reconstructed $B$ meson  candidate, $m_B$ and 
$m_{\rm miss}=\sqrt{(q_{\epem}-\tilde{q}_B)^2}$, where $q_{\epem}$ is the four momentum of
the initial $\epem$ system and $\tilde{q}_B$ is the mass-constrained four momentum of the 
reconstructed $B$ meson candidate.  
By construction, the linear correlation between $m_{\rm miss}$ and $m_B$ vanishes.
Compared to the kinematic variables $\Delta E=E^*_B - \frac 1 2 \sqrt{s}$ and 
$m_{\rm ES}=\sqrt{\frac 1 4 s - p^{*2}_B}$ (where $s=q^2_{\epem}$ and the asterisk
denotes the \epem rest frame), which were used in the previous \babar\ analysis of these modes~\cite{previousbabar}, 
the present combination of variables has less correlation and better background suppression.
The distribution of $m_B$ peaks at the nominal $B$ mass value~\cite{pdg2004},
with a width of about $20\mevcc$ for $\phi\pi^+$, and about $40\mevcc$ for $\phi\pi^0$,
with a low-side tail due to energy leakage from the EMC.
The resolution on $m_{\rm miss}$ is about $5\mevcc$, dominated by the beam-energy spread. 
We require $m_B$ to be within $150~\mevcc$ of the 
nominal $B$  mass and $5.11~\gevcc<m_{\rm miss}<~5.31~\gevcc$. 
The region $m_{\rm miss}<5.2~\gevcc$ is used for background characterization.

The dominant background comes from combinatorial $e^+e^-\to q\bar{q}$ ($q = u,d,s,c$) continuum events.
They tend to be jet-like in the center-of-mass (CM) frame, while \B decays tend to be spherical.
To exploit this characteristic for discriminating against continuum background, we 
use the ratio $L_2/L_0$, where $L_i$ is defined as 
\begin{equation}
L_{i}=\sum_k |p_k||\cos(\theta_k)|^i,
\end{equation}
where  $p_k$ is the momentum of particle $k$, and $\theta_k$ is the angle 
between $p_k$ and the thrust axis 
 of the reconstructed $B$ meson  evaluated in the CM frame. The sum runs over the charged and neutral particles 
of the event not assigned to the $B$ meson.
We require $L_2/L_0<0.55$, which suppresses the continuum background by more than a factor of 3,
while retaining about 90\% of the signal.
We require $|\cos\theta^*_B|<0.9$, where $\theta^*_B$ is the angle between the $B$ candidate 
momentum and the $e^+$ momentum in the CM frame. For $B$ candidates the probability density function of
$\theta^*_B$ is proportional to $\sin^2\theta^*_B$, whereas for continuum events it is 
nearly uniform after acceptance.
We select events for which one $B$ is reconstructed as $B^+ \to \phi h^+$ or $B^0\to \phi \pi^0$
and the other $B$ is only partially reconstructed~\cite{Aubert:2005ja}.
We define $\Delta t$ to be the difference between the proper decay times of the $B$ mesons and 
$\sigma_{\Delta t}$ the uncertainty associated with it. We require, in the case of $B^0\to \phi \pi^0$ only,
$|\Delta t|<20$ \ps and $\sigma_{\Delta t}<2.5$ \ps.
These requirements on $\Delta t$ and $\sigma_{\Delta t}$ retain about 92\% of the signal, 
while removing about 15\% of the continuum events.
The r.m.s. $\Delta t$ resolution is 1.1 \ps~ for the events that
satisfy these requirements.
After the application of these selection criteria on Monte Carlo simulated events, the
efficiencies for $\phi\pi^+$ and $\phi\pi^0$ signal are  $(37.1 \pm 0.1)\%$ and $(29.5 \pm 0.8) \%$ respectively.
The average candidate multiplicity in events with at least one 
candidate is $\sim 1.005$ for both decay modes.
If more than one $B$ candidate is reconstructed in an event, we choose the one 
with the $\phi\to K^+K^-$ invariant mass closest to 
the nominal $\phi$ mass value~\cite{pdg2004}, for $B^+\to \phi\pi^+$ decays. 
For $B^0\to\phi\pi^0$ decays, we choose the candidate with 
the $\pi^0\to\gamma\gamma$ invariant mass closest to the nominal
$\pi^0$ mass value~\cite{pdg2004}. 
These criteria produce no bias in the shape of the other event variables used 
in the maximum likelihood fit described below.
We select $10990$ and  $2732$ events  in the $\phi h^+$ and $\phi\pi^0$  analyses respectively.

A possible background to the 
$\phi \rightarrow K^+K^-$ decays comes from the S-wave production of the 
$K^+K^-$ system ($B\to (K^+K^-)_{\rm S-wave}\pi$ decays) with contributions coming predominantly from resonances 
such as $f_0(980)$ and $a_0(980)$.
Using samples of simulated decays of $B$ mesons equivalent to 
nearly five times the size of the data sample, we found that all the other $B$ decay modes
give negligible sources of background.
To discriminate against S-wave background in the maximum likelihood fit, we use 
the helicity of the $K^+K^-$ system, in terms of 
the cosine of the  angle $\theta_H$ between the $K^+$ candidate
and the parent $B$ meson flight direction in the $K^+K^-$ rest frame. 
The helicity probability density function is proportional to $\cos^2\theta_H$ for the signal, 
and is uniformly distributed for the S-wave background.
Further discrimination is provided by the $K^+K^-$ invariant mass distribution, $m_{KK}$, which 
peaks at the $\phi$ mass for the signal, while it peaks at lower values for the 
S-wave background.

In the case of charged $B$ decays, we exploit the Cherenkov angle $\theta_c$ measured in the DIRC for the primary track,
in order to determine simultaneously the yields of $B^+ \to \phi \pi^+$ and $B^+ \to \phi K^+$ decays 
and the yields of the two corresponding $B^+\to (K^+K^-)_{\rm S-wave}h^+$ ($h=\pi,K$) background components.

\label{sec:fit}
Signal and background yields $N_i$, where $i$ denotes signal, continuum,
and S-wave background,
are extracted using an extended maximum likelihood 
fit with the likelihood function:

\begin{equation}
\label{eqn:like}
{\mathcal L} = \frac{1}{N!}\exp{\left(-\sum_{i}N_{i}\right)}
\prod_{j=1}^N\left[\sum_{i}N_{i}{\mathcal P}_{i}(\vec{x_j};\vec{\alpha}_i)\right]\!,
\end{equation}
where $N$ is the total number of events entering the fit.
The probabilities ${\mathcal P}_{i}$ 
are products of Probability Density Functions (PDF) for each of the independent variables 
$\vec{x} = \left\{m_{\rm miss}, m_B, {L_2/L_0}, m_{KK}, \cos\theta_H \right\}$. 
In the case of $B^+ \to \phi h^+$ the variable $\theta_c$ is also used in the fit.
The $\vec{\alpha}_i$ are the parameters of the PDFs for $\vec{x}$.
The continuum parameters are allowed to vary, except for the 
$m_{\rm miss}$ end-point. 
All other parameters $\alpha_i$ are fixed to their values derived from data
 control samples.
These are varied within their 
uncertainties to evaluate the systematic error.
By minimizing the quantity $-\ln{\mathcal L}$ in two separate fits, 
we determine the yields  for $\phi \pi^+$ and 
$\phi \pi^0$.
There are three $B$ backgrounds to $B^+\to\phi \pi^+$ decay ($B^+\to (K^+K^-)_{\rm S-wave}h^+$ 
and $B^+\to\phi K^+$), while only $B^0\to (K^+K^-)_{\rm S-wave}\piz$ contributes to the $B^0\to\phi\piz$ mode.
All the yields in Eq.~\ref{eqn:like} are allowed to fluctuate to negative values in the fits.

The distributions of $L_2/L_0$ and $\cos\theta_H$ are described by a 
parametric step function \cite{Aubert:2003hf} and a second-order polynomial 
respectively. We use a 
Gaussian for the $m_{\rm miss}$ distribution for $\phi\pi^+$ signal and S-wave components, 
a Gaussian with  exponential tails for the  $m_B$ distribution for $\phi\pi^+$ signal and S-wave components,
and for both $m_{\rm miss}$ and $m_B$ for $\phi\pi^0$ signal and S-wave components.
For the continuum $m_{\rm miss}$ distribution we use the function $x\sqrt{1-x^2}\exp{\left[-\xi(1-x^2)\right]}$,
with $x\equiv2m_{\rm miss}/ \sqrt{s}$ and $\xi$  a floating parameter. 
The $m_{KK}$ invariant mass distribution is 
described by a relativistic Breit-Wigner function for signal, a relativistic Breit-Wigner plus 
exponential for the continuum background and 
a Flatt\'{e} 
\cite{flatte,Baru:2004xg} function for the  S-wave background.
The Flatt\'{e} function takes into account the coupling of the scalar 
resonances to the $\pipi$ and $\Kp\Km$ channels~\cite{SHEN:2004sw}.

The Cherenkov angle $\theta_c$ PDFs are obtained from a large data sample 
of  $D^{*+}\to D^0\pi^+$ ($D^0\to K^-\pi^+$) decays 
where $K^\mp/ \pi^\pm$ tracks are identified through the charge correlation with the $\pi^\pm$ from the 
$D^{*\pm}$ decay. The PDFs are constructed separately for $K^+$, $K^-$, $\pi^+$ and $\pi^-$ tracks as a function of
momentum and polar angle using the expected values of~ $\theta_c$, and its uncertainty.

Using a large number of simulated experiments, we find that 
the usual maximum likelihood fitting technique does not provide an 
unbiased  estimate of the true values of signal and S-wave yields 
($N_{S}$ and $N_{\rm S-wave}$) because of the non-Gaussian shape of the likelihood function
when the yield is very small.
Therefore we use a Bayesian statistical approach
to obtain a modified likelihood function $L(N_S)$:
\begin{equation}
L(N_S)=N_0\int_0^\infty dN_{\rm S-wave}{\mathcal L}(N_S,N_{\rm S-wave}),
\label{eq:like1dim}
\end{equation}
where the normalization $N_0$ is such that $\int_0^\infty dN_S 
L(N_S)=1$. 
The two dimensional likelihood ${\mathcal L}(N_S,N_{\rm S-wave})$ is given
at each point on the $N_S$-$N_{\rm S-wave}$ plane by the function defined in 
Eq.~\ref{eqn:like}, maximized with respect to all of the other fit variables.
When seeking the central value for the branching fraction we take the median
of $L$, with the lower limit replaced by $-\infty$ and $N_S$ unrestricted.
This is because we find from simulations that in the case of very low
yields, the median provides a less biased estimator of the true value of
$N_S$ than the maximum of $L$.  
We correct the central value of the branching fractions for the residual biases.
When calculating upper limits, we impose
the {\it a priori} constraints $N_S > 0$ and $N_{\rm S-wave} > 0$.

\begin{figure*}[!tb]
\begin{center}
\epsfig{file=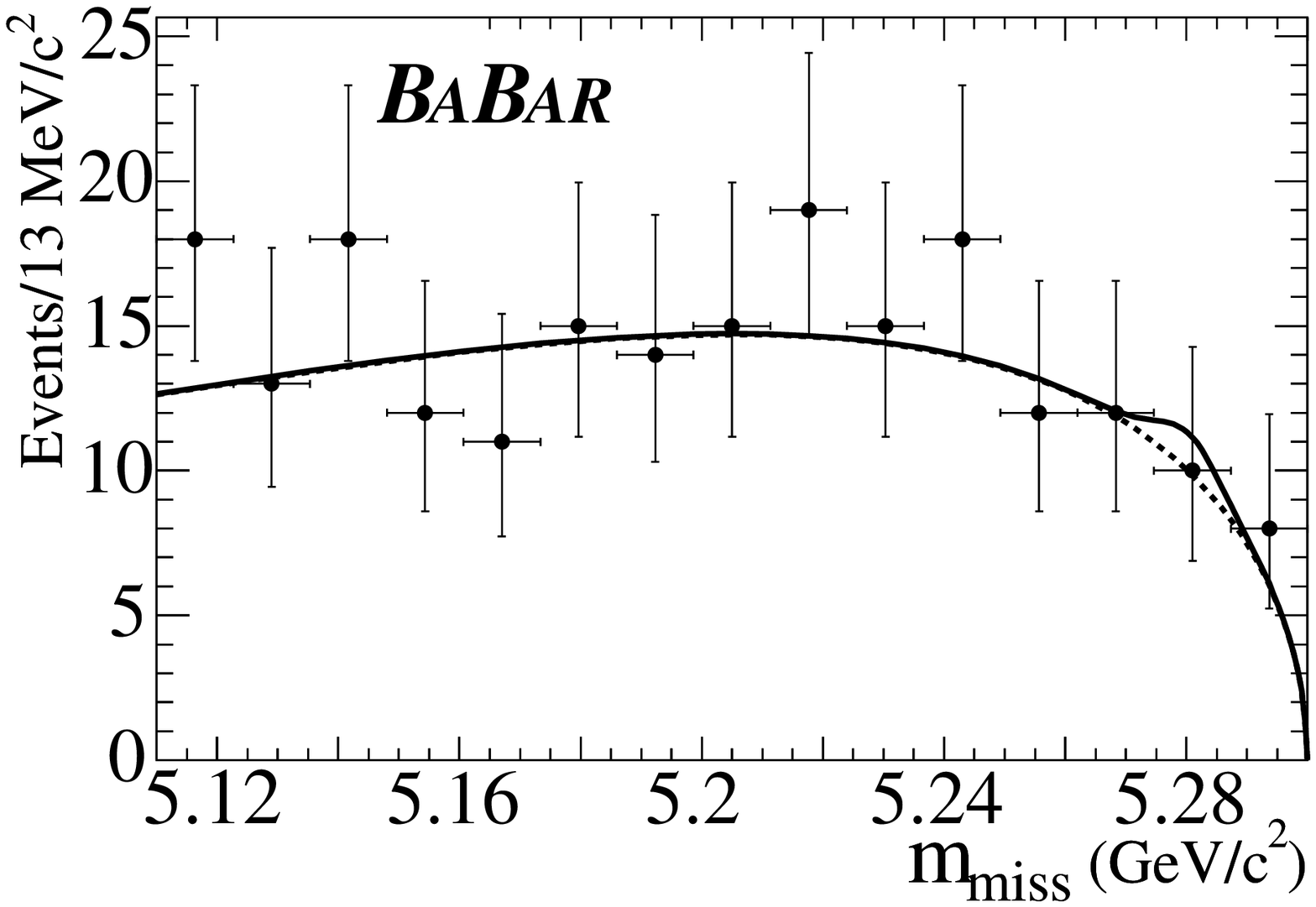,width=5.9cm}
\epsfig{file=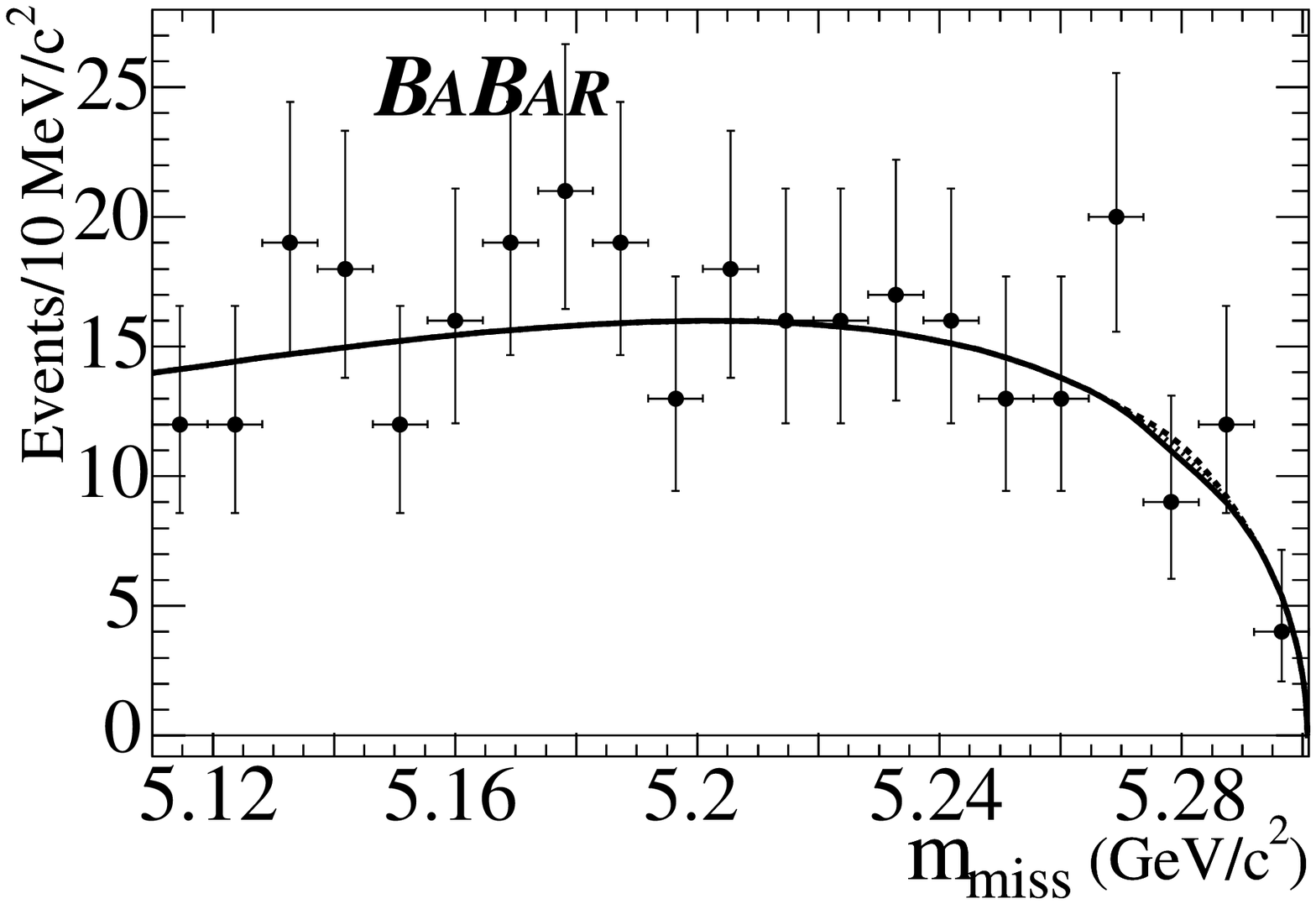,width=5.9cm}
\epsfig{file=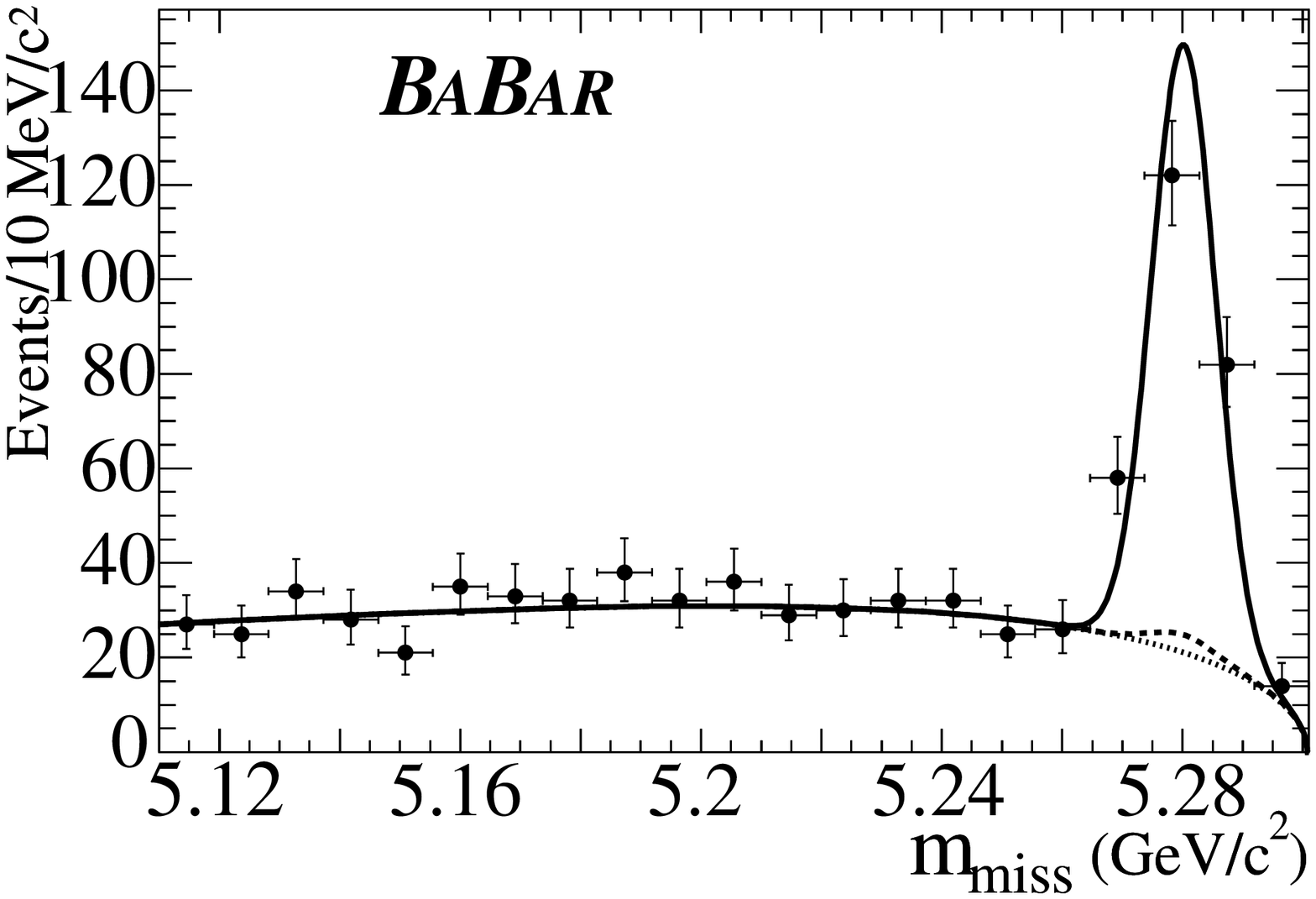,width=5.9cm}
\epsfig{file=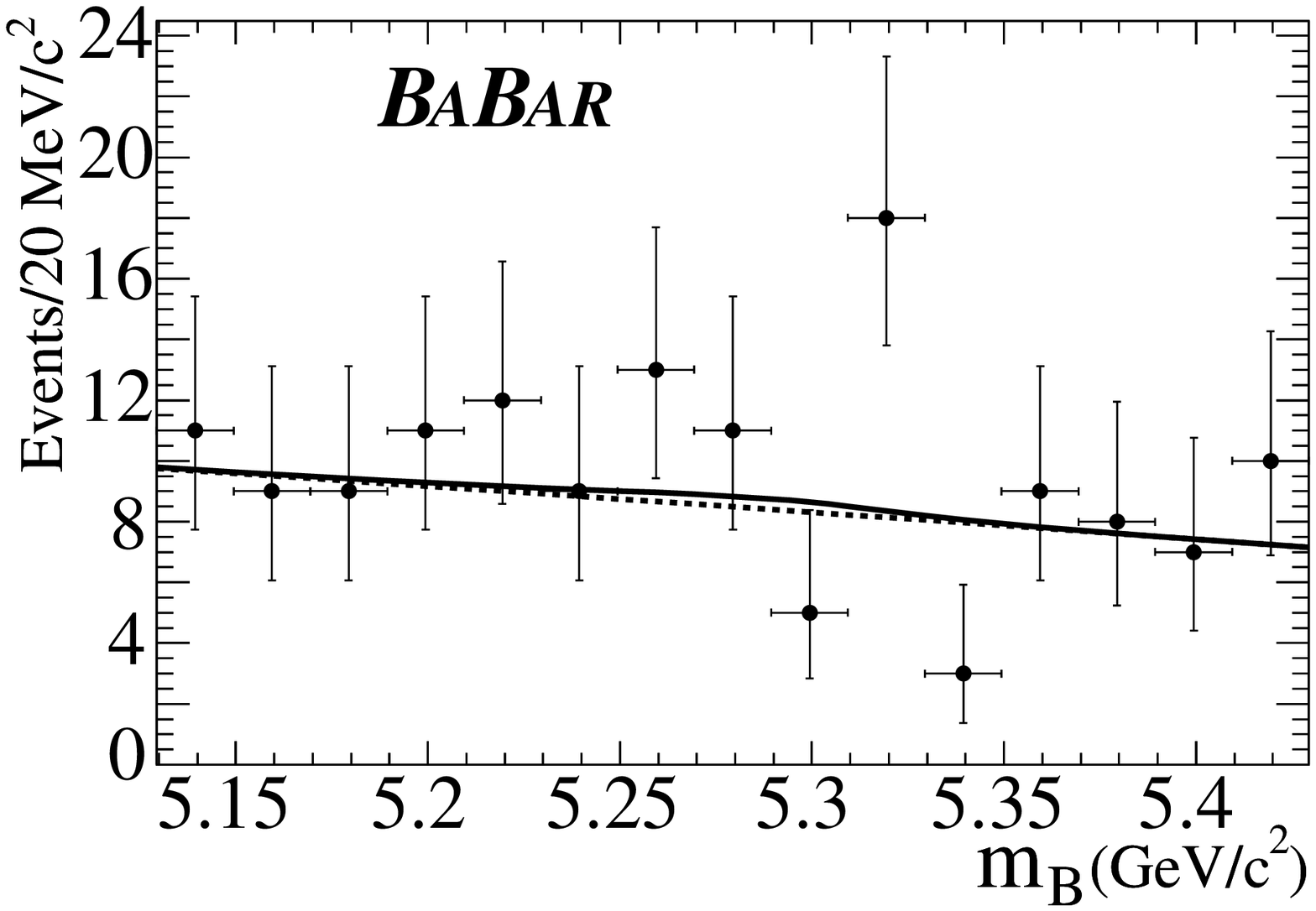,width=5.9cm}
\epsfig{file=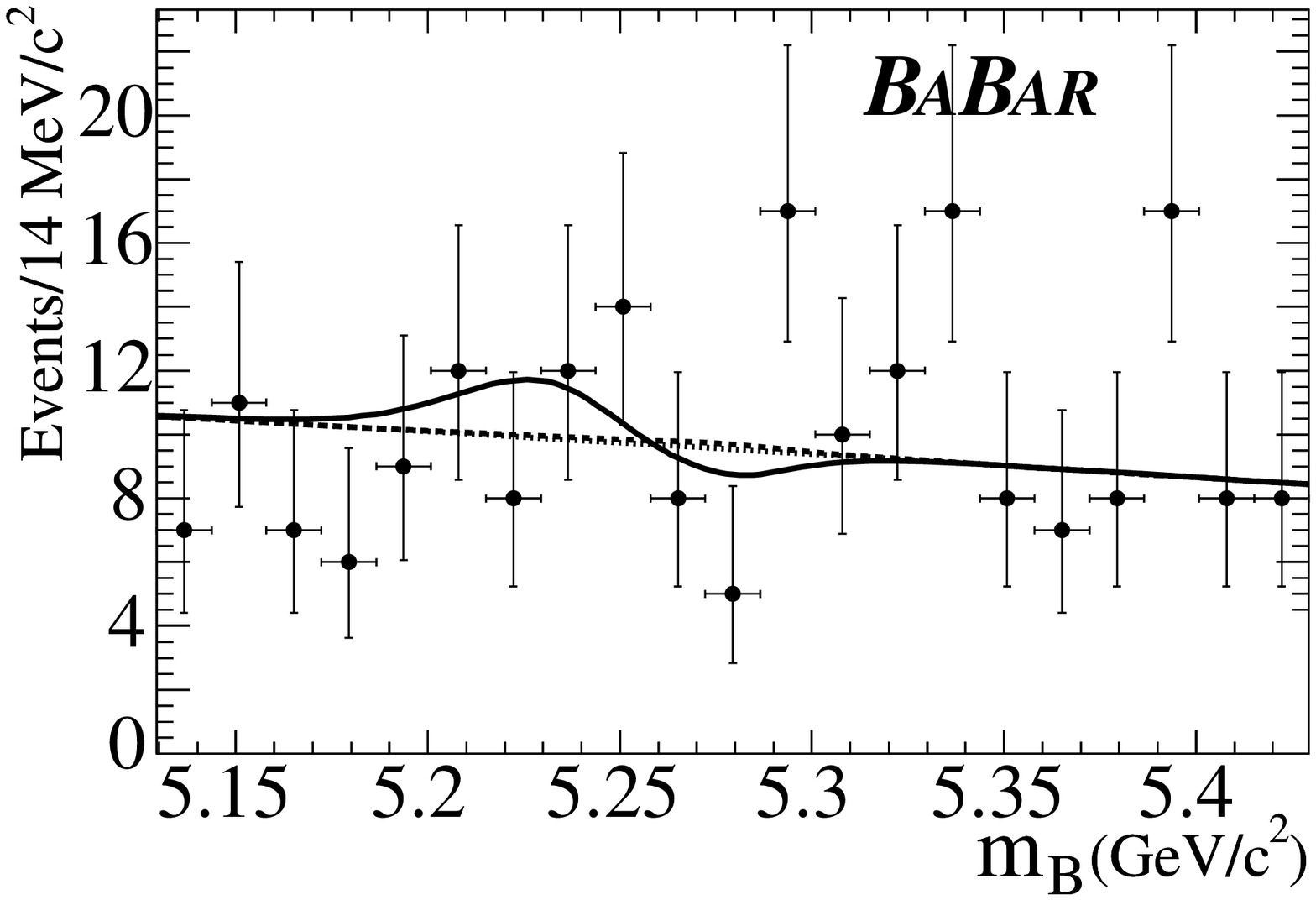,width=5.9cm}
\epsfig{file=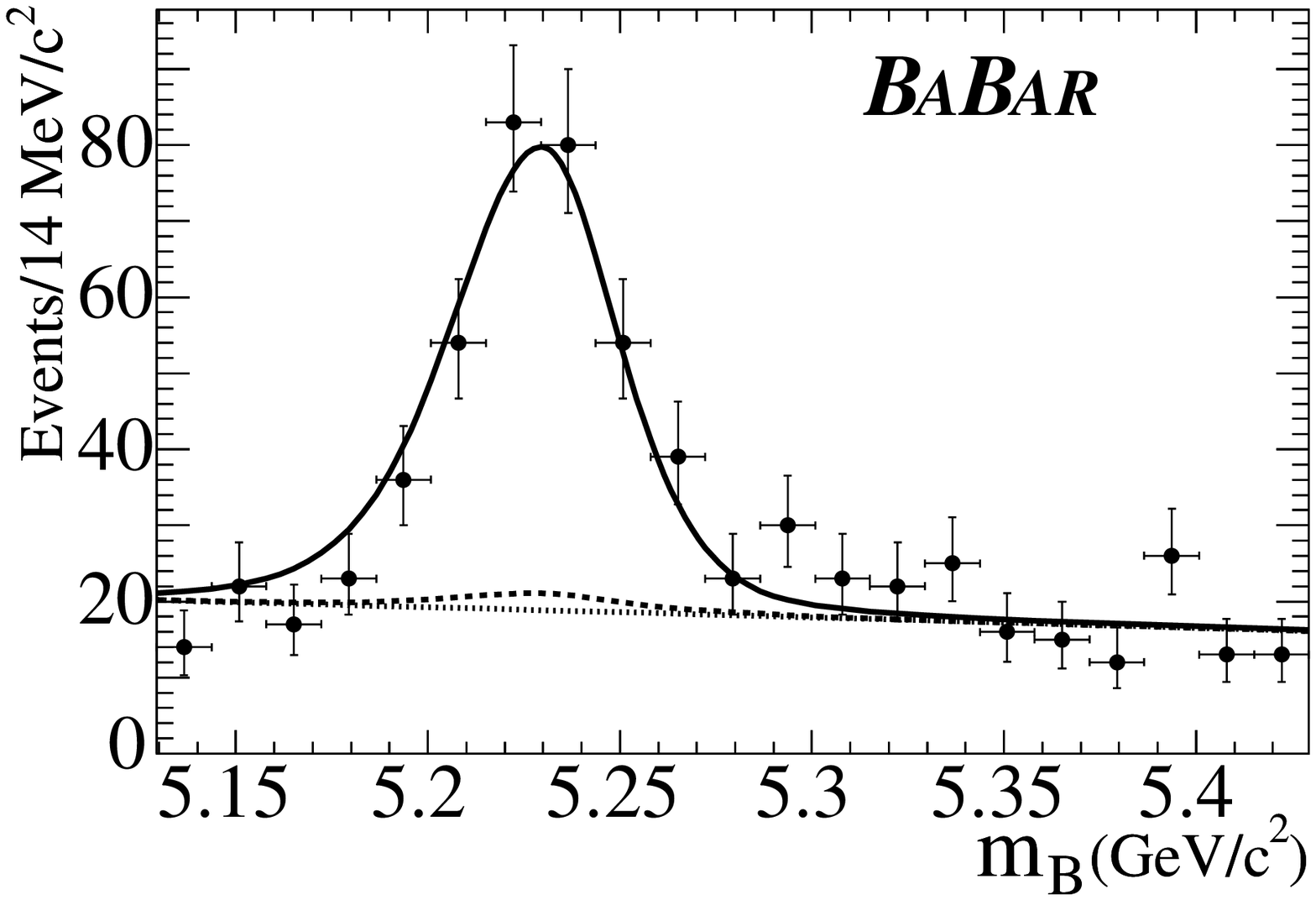,width=5.9cm}
\caption{\label{fig:projplots}Distribution of $m_{\rm miss}$ (top) and $m_{B}$ (bottom) 
for reconstructed $\B^0\to \phi\pi^0$ (left), $B^+\to \phi\pi^+$ (middle)
and $B^+ \to \phi h^+$ ($h=\{\pi,K\}$) (right), after applying a requirement on 
the ratio of signal likelihood to signal-plus-background likelihood to enhance the signal.
The curves are projections from the likelihood fit for total yield (solid line), for the 
continuum background (fine dashed line) and for continuum plus S-wave component (dashed line).
For $B^+\to\phi h^+$ decay we do not apply any particle identification criteria
and assign the pion mass to the primary track. For this reason, the $m_B$ distribution 
for $B^+\to\phi K^+$ events is shifted with respect to the nominal $B^+$ mass
(positive bump in bottom middle and bottom right plot), while it peaks at the nominal value for
$B^+\to\phi \pi^+$ events (negative bump in bottom middle plot).
}
\end{center}
\end{figure*}

Fig.~\ref{fig:projplots} shows the $m_{\rm miss}$ and $m_B$ 
distributions for data, with the PDF corresponding to the maximum 
likelihood fit overlaid.
We do not observe evidence for either $\Bp\to\phi\pip$ or $B^0 \to 
\phi\piz$ decays.

\begin{table}[!ht]
\caption[.]{ \label{tab:summ} Signal yield (evaluated as the median of the likelihood), detection efficiency $\varepsilon$ (the uncertainty
includes both statistical and systematic effects),
measured branching fraction $\cal B$ with statistical error, after the correction for the fit bias has been applied, 
for the two decay modes considered and upper limit at 90\% probability.}
\begin{center}
\setlength{\tabcolsep}{0.3pc}
\begin{tabular}{lccc}
\hline\hline
&& \\[-0.3cm]
              &   $B^+ \to \phi\pi^+$  &    $B^0 \to \phi\pi^0$        \\
\hline
	Yield			& $-1.5 \pm  5.9$  & $~~4.0 \pm  3.5$ \\ 
	$\varepsilon(\%)$	& $37.1 \pm 0.1$ &  $29.5 \pm 0.8$\\
	${\cal B} (10^{-6})$	& $-0.04 \pm 0.17$& $~~0.12 \pm 0.13$  \\
	$\mathrm{UL}({\cal B}) (10^{-7})$& $2.4$&   $2.8$\\
\hline\hline
\end{tabular}
\end{center}
\vspace{-0.5cm}
\end{table}

The signal yields, extracted from the median of the likelihood $L(N_S)$ (Eq. ~\ref{eq:like1dim}), 
are reported in Table~\ref{tab:summ}. In the case of $B^+ \to \phi h^+$, we also measure the $N_{\phi K^+}$ yield, 
which is found to be compatible with the expectation from published branching fractions \cite{bfphik}.

The branching fraction {\BR} is calculated from the observed number of 
signal events as 
\begin{equation}
{\cal B} = \frac{N_{S}}{\varepsilon \cdot N_{B\bar B} \cdot \BR(\phi\to \Kp\Km)}
\label{eq:br}
\end{equation}
where $N_{B\bar B}$ is the number of 
$B\bar B$ pairs produced and $\varepsilon$ is the reconstruction 
efficiency for the $B$ candidates. In Eq.~\ref{eq:br} we assume
equal branching fractions for $\Upsilon(4S)$ decays to charged and 
neutral $B$-meson pairs~\cite{upsbf}.

The systematic uncertainties are summarized in Table~\ref{tab:syst}.
\begin{table}[bt]
\caption[.]{ \label{tab:syst} Summary of systematic uncertainties contributing to the 
total error for the upper limit on the branching fraction. They are given in units of $10^{-8}$.}
\begin{center}
\setlength{\tabcolsep}{0.5pc}
\begin{tabular}{lcc}
\hline\hline
&& \\[-0.3cm]
							&  $\Bp\to\phi\pi^+$ &     $\Bz\to\phi\pi^0$  \\
\hline												    
&& \\[-0.3cm]											    
PDF Uncertainty					&       $^{+1.9}_{-2.8}$		    &       $^{+3.6}_{-4.2}$	\\
&& \\[-0.3cm]											    
PID Efficiency					&       $0.1$				    &       $0.1$			\\
Tracking Efficiency			&       $0.1$				    &       $0.2$			\\
$\piz$ Efficiency				&       - 				    &       $0.1$			\\
$L_2/L_0$ Cut				&       $0.1$				    &       $0.3$			\\
$B\bar B$ Pair Counting				&       $0.1$				    &       $0.2$			\\
Interference Effects			&       $0.3$				    &       $0.6$			\\
&& \\[-0.3cm]											    
${\cal B}(\phi\to\KpKm)$, ${\cal B}(\piz\to\gamma\gamma)$		&       $0.1$				    &       $0.1$			\\
\hline												    
&& \\[-0.3cm]											    
Total					&       $^{+2.8}_{-3.6}$		    &       $^{+3.7}_{-4.3}$		\\
\hline\hline
\end{tabular}
\end{center}
\vspace{-0.5cm}
\end{table}
The uncertainty arising from the lack of knowledge of continuum background 
PDFs is part of the statistical error since the background parameters are free to vary in the fit.
The uncertainty on the  signal PDFs represents the dominant error.
We estimate it by using simulated and high-statistics data control samples of
$\Bu \to \pip \Dzb \ (\Dzb \to \Kp \pim)$
and $\Bz \to \pip \Dm \ (\Dm \to \KS \pim)$ events. 
In order to estimate the systematic uncertainty on $m_{B}$ for  $B^0 \to \phi \pi^0$
we use a data control sample of $B^+ \to h^+\piz$ events.
The control channels have event topologies similar
to those of $\Bu \to \phi h^+$ and $\Bz \to \phi\pi^0$.
We use them  to determine 
the signal PDF parameters and take  the difference
in yields found by varying these parameters within one standard deviation 
as the systematic error.
The second most important error comes from the 
uncertainty on the efficiency $\varepsilon$.  
The track detection efficiency uncertainty is estimated to be 0.8\%
per track from a study of a variety of control samples, such as $\tau\to 3$-track decays.
We assign 0.5\% uncertainty on the kaon identification efficiency.
The uncertainty on the reconstruction efficiency for the $\piz$ is 3\%, as measured in
a large sample of $\tau^- \to \rho^-\nu_\tau$, $\rho^-\to \pi^-\piz$ decays coming from  $\epem \to \tau^+\tau^-$.
We assign a 1.8\% uncertainty on the  $L_2/L_0$ 
cut efficiency, estimated by the difference between Monte Carlo and data control samples,
1.1\% on the total number of $\Upsilon(4S) \to 
\B\bar B$ decays in the sample 
 and 1.2\% on the knowledge of 
$\BR(\piz \to \gamma\gamma)$ and $\BR(\phi \to \Kp\Km)$.
We estimate the systematic error introduced by the 
approximation of ignoring interference effects between the $\phi$ and 
the $\Kp\Km$ S-wave components by
varying  the relative strong phases  and taking  the 
largest observed variation as the 
error.  In this study we include the $f_0(980)$ resonance and a non-resonant 
component,  whose contribution is taken from a $\B^+ \to 
\Kp\Km\Kp$ Dalitz plot measurement by the Belle Collaboration\cite{BelleDalitz}.
The resulting uncertainty is 4.4\% for both modes.

\begin{figure}[!tb]
\begin{center}
\epsfig{file=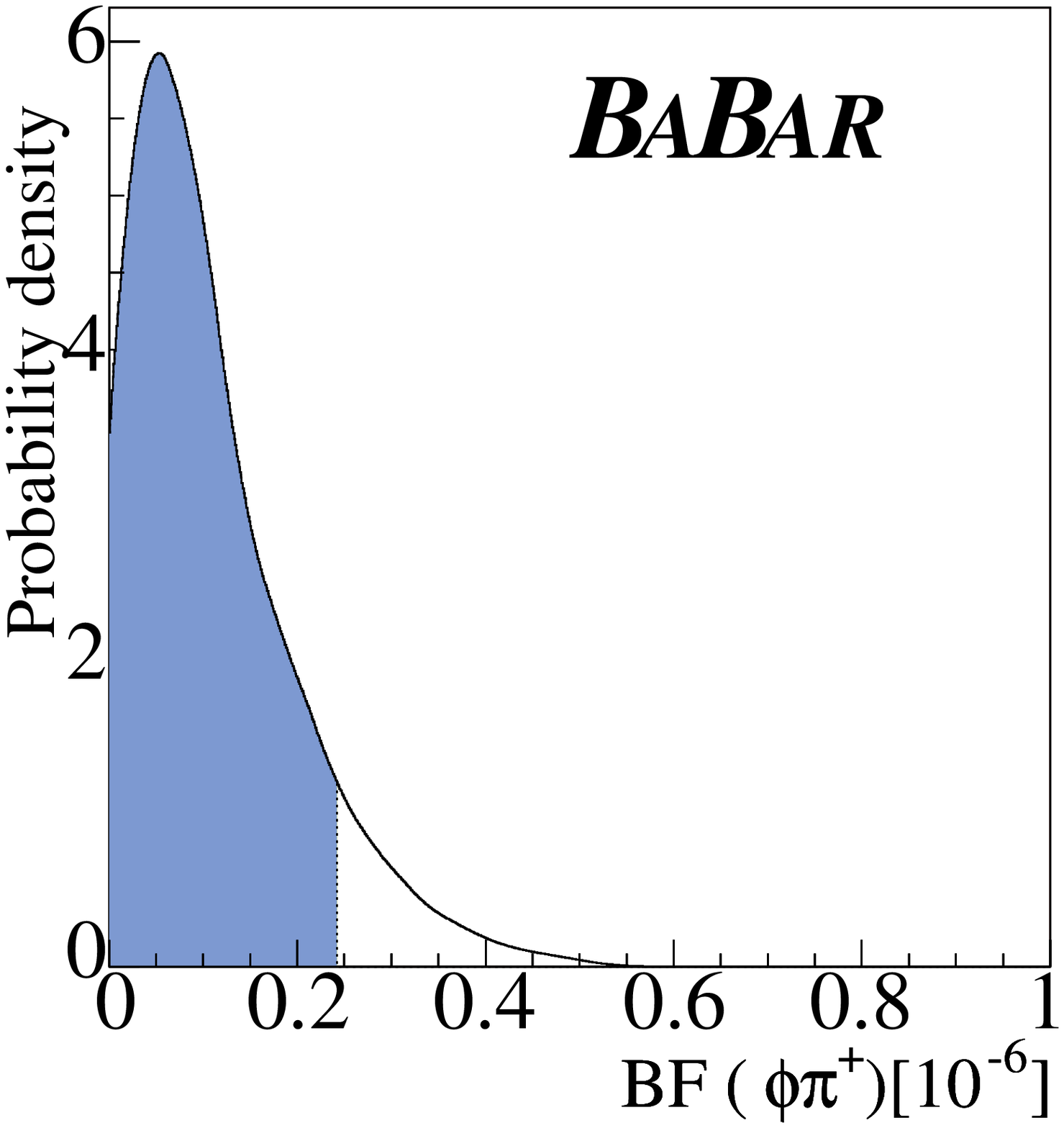,width=4.25cm}
\epsfig{file=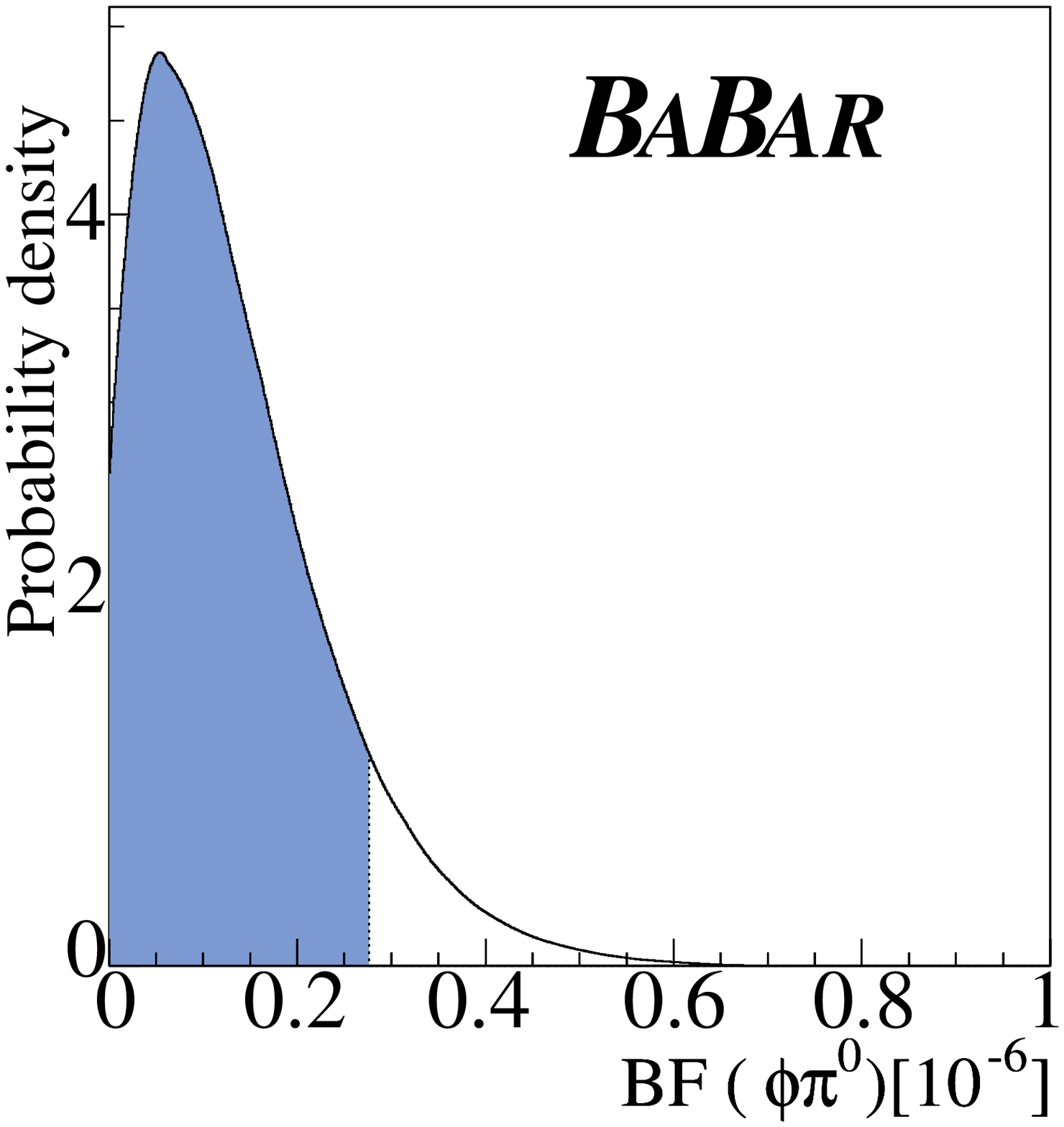,width=4.25cm}
\caption{\label{fig:lik}Likelihood distribution, $L_{\cal B}({\cal B})$, for ${\cal B}(B^+ \to \phi \pi^+)$ (left)
and ${\cal B}(\Bz \to \phi \piz)$ (right) in arbitrary units. The upper boundary of the dark region represents the $90\%$ probability 
upper limit.}
\end{center}
\end{figure}

Under the assumption that $N_{B\bar B}$ and $\varepsilon$ are 
distributed as Gaussians, 
we obtain a likelihood function, $L_{\cal B}$, for the branching fraction, ${\cal B}$,
based on Eq.~\ref{eq:br}, by
convolving the likelihood ($L$ in Eq.~\ref{eq:like1dim}) 
 with the distributions of $N_{B\bar B}$ and $\varepsilon$. We also 
include the additional
uncertainty coming from the systematic error on  the signal yield.
The resulting likelihood is shown in Fig.~\ref{fig:lik} for each of the two decay modes.
In the plots, the upper boundary of the dark region represents the $90\%$ probability 
Bayesian upper limit ${\cal B}_{\mathrm{UL}}$, defined as:
\begin{equation}
\int_{0}^{{\cal B}_{\mathrm{UL}}}{L_{\cal B}}({\cal B}) d{\cal B} ~=~ 
\frac {9} {10}\int_{0}^{+\infty}{L_{\cal B}}({\cal B}) d{\cal B}
\end{equation}
 We determine ${\cal B}(\B^+ \to \phi\pi^+)<2.4 \times 10^{-7}$ and ${\cal 
B}(B^0 \to \phi\pi^0)<2.8 \times 10^{-7}$.

We compute the central values for the branching fractions by
correcting the fitted signal yields for the fit bias,
estimated using a large number of simulated experiments, and including 
a systematic uncertainty equivalent to half the fit bias.
This error corresponds to a shift of
$-0.8$ and $-0.4$ events in the signal yield, and  $-1.9 \times 10^{-8}$ 
and $-1.2 \times 10^{-8}$ in the branching fraction
for $B^+\to \phi\pi^+$ and $\Bz\to\phi\piz$, respectively. 
Without taking into account the \textit{a priori} knowledge of $N_S>0$ and 
$N_{\rm S-wave}>0$, and
integrating the likelihood in Eq.~\ref{eq:like1dim} around the median, we  obtain 
as $68\%$-probability regions
${\cal B}(B^+\to\phi\pi^+)=\left(-0.04 \pm 0.17  \right) \times 10^{-6}$ and
${\cal B}(B^0 \to \phi\pi^0)= \left(0.12 \pm 0.13 \right) \times 10^{-6}$.
The results are summarized in Table~\ref{tab:summ}.

In summary we have searched for 
$B^+ \to \phi \pi^+$ and $B^0 \to \phi \pi^0$ decays
 in a sample of 232 million \BB meson pairs. We find no
 evidence of signal  and therefore we place
upper limits  ${\cal B}(\B^+ \to \phi\pi^+)<2.4 \times 10^{-7}$ and 
 ${\cal B}(B^0 \to \phi\pi^0)<2.8 \times 10^{-7}$ at 90\% probability. 
These limits are more stringent than earlier
results~\cite{bfphik,previousbabar,cleo} 
and they supersede our previous publications~\cite{bfphik,previousbabar}.
They are consistent with existing SM predictions~\cite{Bar-Shalom:2002sv}.

\input{pubboard/acknow_PRL.tex}

\end{document}

%% file: pubboard/authors_feb2006.tex
%
\author{B.~Aubert}
\author{R.~Barate}
\author{M.~Bona}
\author{D.~Boutigny}
\author{F.~Couderc}
\author{Y.~Karyotakis}
\author{J.~P.~Lees}
\author{V.~Poireau}
\author{V.~Tisserand}
\author{A.~Zghiche}
\affiliation{Laboratoire de Physique des Particules, F-74941 Annecy-le-Vieux, France }
% ...
\collaboration{The \babar\ Collaboration}
\noaffiliation

%% file: pubboard/acknow_PRL.tex
We are grateful for the excellent luminosity and machine conditions
provided by our \pep2\ colleagues, 
and for the substantial dedicated effort from
the computing organizations that support \babar.
The collaborating institutions wish to thank 
SLAC for its support and kind hospitality. 
This work is supported by
DOE
and NSF (USA),
NSERC (Canada),
IHEP (China),
CEA and
CNRS-IN2P3
(France),
BMBF and DFG
(Germany),
INFN (Italy),
FOM (The Netherlands),
NFR (Norway),
MIST (Russia), and
PPARC (United Kingdom). 
Individuals have received support from the
Marie Curie EIF (European Union) and
the A.~P.~Sloan Foundation.

%% file: phipi.bbl
\begin{thebibliography}{99}


\bibitem{Bar-Shalom:2002sv}
S.~Bar-Shalom, G.~Eilam and Y.~D.~Yang,
Phys.\ Rev.\ D {\bf 67}, 014007 (2003).

\bibitem{charge}
Throughout this paper, 
charge conjugate reactions are included implicitly and 
 $\phi$ refers to the $\phi(1020)$.

\bibitem{burassilv} See for example   A.~J.~Buras and L.~Silvestrini,
  Nucl.\ Phys.\ B {\bf 569} 3 (2000).

\bibitem{gqbound}
Y.~Grossman {\it et al.},
Phys.\ Rev.\ D {\bf 68}, 015004 (2003).

\bibitem{bfphik} 
\babar{} Collaboration,  B.~Aubert {\it et al.},
  Phys.\ Rev.\ D {\bf 69} 011102 (2004).

\bibitem{previousbabar}
\babar{} Collaboration,  B.~Aubert {\it et al.},
  Phys.\ Rev.\ D {\bf 70} 032006 (2004).


\bibitem{cleo} CLEO Collaboration, T.~Bergfeld {\it et al.}, \jprl{81}, 272 (1998).

\bibitem{ref:babar}
\babar\ Collaboration, B. Aubert {\em et al.}, \nima{479}, 1 (2002).

\bibitem{pep}
PEP-II Conceptual Design Report, SLAC-R-418 (1993).

\bibitem{Lange:2001uf}
  D.~J.~Lange,
  \nima{462}, 152 (2001).

\bibitem{geant}
GEANT Collaboration,  S.~Agostinelli {\it et al.}, 
  \nima{506}, 250 (2003).

\bibitem{pdg2004}
Particle Data Group, S.~Eidelman {\it et~al.} Phys.~Lett.~B {\bf 592}, 1 (2004).

\bibitem{Aubert:2005ja}
  \babar{} Collaboration, B.~Aubert {\it et al.}
  Phys.\ Rev.\ D {\bf 71}, 091102 (2005)

\bibitem{Aubert:2003hf}
\babar{} Collaboration,  B.~Aubert {\it et al.},
  Phys.\ Rev.\ Lett.\  {\bf 91}, 241801 (2003).

\bibitem{flatte} S.~Flatt\'{e}, Phys.\ Lett.\ B {\bf 63}, 224 (1976).
\bibitem{Baru:2004xg}
  V.~Baru, J.~Haidenbauer, C.~Hanhart, A.~Kudryavtsev and U.~G.~Meissner,
  Eur.\ Phys.\ J.\ A {\bf 23}, 523 (2005).

\bibitem{SHEN:2004sw}
  X.~Shen,
  Int.\ J.\ Mod.\ Phys.\ A {\bf 20}  1706 (2005).


\bibitem{upsbf} \babar{} Collaboration, B.~Aubert {\it et al.},
  Phys.\ Rev.\ D\  {\bf 69}, 071101 (2004).

\bibitem{BelleDalitz}
Belle Collaboration, A.~Garmash {\it et al.}, \jprd{71}, 092003 (2005).

\end{thebibliography}
